\documentclass[twocolumn,aps,floatfix,superscriptaddress]{revtex4}

\usepackage{bm}
\usepackage{graphicx}
\usepackage{amssymb}
\usepackage{amsmath}
\usepackage{hyperref}

\newcommand{\la}{\label}
\newcommand{\beq}{\begin{equation}}
\newcommand{\eeq}{\end{equation}}
\newcommand{\be}{\begin{equation}}
\newcommand{\ee}{\end{equation}}

\newcommand{\p}{\partial}

\begin{document}
\title{Braid group representations and cold Fermi gases in the fast pairing regime} 

\author{Bryce Hotalen}
\address{Dept. of Electrical Engineering, University of South Florida and TriQuint Semiconductor, Inc.}

\author{Razvan Teodorescu}
\address{Dept. of Mathematics and Statistics, University of South Florida}
\email{razvan@usf.edu}

\begin{abstract}
It is widely recognized that the main difficulty in designing devices which could process information using quantum states is due to the decoherence of local excitations about a ground state. A solution to this problem was suggested in \cite{Kitaev}, relying on (non-local) topological excitations,   structurally protected against local noise. However, a practical implementation of this proposal using special Landau levels in fractional quantum Hall effect systems (FQHE) \cite{QHE} has proven elusive, while accessible FQHE states are theoretically not optimal because their representations in the Hilbert space of states are not dense. We propose using a different physical system (cold Fermi atoms), whose semiclassical dynamics is described by a hyperelliptic function in the Sklyanin formalism. The homological structure of the complex curve corresponds to representations of the braid group, with the action of Hecke operators leading to singularities detectable in the semiclassical oscillations. We argue that, for a fixed genus of the hyperelliptic curve,  the Richardson-Gaudin pairing Hamiltonian problem is the singular limit $m \to \infty$ of level-$k$ $\widehat{sl}(2)$, with $k+2 = \frac{4}{8m+1}\to 0$, so that the level $k$ is admissible in the sense of Kac and Kazhdan \cite{KK}, but the corresponding Hecke algebra is a $q-$deformation of the symmetric group with fixed $q = e^{i\pi/4}$, as $m \to \infty$.  
\end{abstract}


\maketitle



\paragraph*{1. Introduction}

Two decades ago, the introduction of the first quantum algorithm \cite{Shor} led to an intense activity aimed at finding physical systems whose quantum states could be used in a reliable way to process information. In a quantum system, information is encoded in quantum bits (or qubits), rather than a classical bit. Unlike a classical bit, which can only hold a value of 0 or 1, a qubit can be 0, 1, or any irreducible superposition of both. A qubit is realized by way of a two state quantum system. A popular approach to development of a qubit has been to make use of Rabi oscilations. Rabi oscilations are an external induced sinusoidal time evolution of a quantum system between its two states, typically by way of optical or electromagnetic stimulation.  This behavior has been fundamental to many approaches in realization of a working quantum computer.  The Rabi oscillations can be geometrically represented through use of the Bloch sphere. 
The classical states of 0 and 1 correspond to the poles of the sphere. As the system evolves in time the state vector oscilates between the poles  of the sphere by a variety of paths. 
 However, in spite of experimental successes designing various types of qubits 
\cite{qubits}, a successful implementation of quantum algorithms is still elusive, mainly because of the fundamental effect of decoherence which characterizes all devices based on local quantum excitations. Decoherence is the irreversible loss of a quantum systems ability to process information; it arises when a quantum system becomes coupled to its environment (which is unavoidable if we wish to control the evolution of the system). This coupling results in the information previously contained in the system being lost to the environment rendering further computation impossible. Decoherence is represented on the Rabi sphere by the state vector rotating around the equator of the sphere, in essence being neither 0 nor 1 nor a superposition.  
This effect, studied under various assumptions on the type of noise \cite{noise} describing the coupling of the system to an external bath, is very important because, for devices based on local excitations (or quantum oscillations associated with single-state quantum dynamics), noise and useful signals are indistinguishable. To address this important issue, A. A. Kitaev and collaborators \cite{Kitaev} have proposed using non-local, topological excitations, which are basically unaffected by most types of noise from coupling with the environment. An essential ingredient in the construction of these special excitations is the fact that they require the simultaneous creation (or annihilation) of two excitations at distinct points $z_{1}, z_{2}$. As explained in \cite{Kitaev}, the main advantage of working with such states comes from their {\emph{non-locality}}: the likelihood that noise would create two excitations simultaneously at $z_1, z_2$ is completely negligible, so that the system is protected against a large class of perturbations, namely local random fluctuations.

%

%
%


\paragraph*{2. Topological excitations from Quantum Hall states} 

The specific physical model which seemed to allow a direct implementation of Kitaev's proposal is the Fractional Quantum Hall Effect (FQHE), more precisely the partially filled Landau level at filling factor 
$\nu = 5/2$. Following \cite{52level}, we briefly recall the essence of this theory:  
\begin{itemize}
\item[(i)] the ground state of the $\nu=\frac{5}{2}$-filling fraction FQHE Landau level is given by the Pfaffian construction
\be
\Psi_{GS} (z_1, \ldots, z_{2n})\sim {\rm{Pf}}\left [\frac{1}{z_i-z_j} \right ] \prod_{k<l}(z_k-z_l)^2,
\ee 
(ignoring single-particle factors), while an anyon excitation (obtained by removing two particles, i.e. by creating two {\emph{holes}} at $z_{1,2}$) corresponds to
$$
\Psi_{2h} (z_1, z_2; z_3, \ldots, z_{2n})\sim  {\rm{Pf}}\left [\frac{z_1-z_2}{z_1z_2(z_i-z_j)} \right ] \prod_{k<l}(z_k-z_l)^2,
$$
up to single-particle factors, where Pf stands for the Pfaffian of an even-dimensional, skew-symmetric matrix, ${\rm{Pf}}(A) = \sqrt{{\rm{Det}}(A)}$; 
\item[(ii)] the  two-hole excitation changes under a permutation of its arguments as 
$$
\Psi_{2h}(z_2, z_1; z_3, \ldots, z_{2n}) = e^{\frac{2 i \pi}{\kappa}} \Psi_{2h}(z_1, z_2; z_3, \ldots, z_{2n}),
$$
obeying the anyon statistics with $\kappa = 4$ ($\kappa=1,2$ correspond to bosons and fermions, respectively).  
\end{itemize}

Unfortunately, another feature of this theory makes it {\emph{theoretically not optimal}} for achieving quantum computation, and prompts current research for finding similar non-commutative anyon theories: specifically, it belongs to a class of theories whose states form irreducible representations for the $\widehat{sl}(2)_k$ Wess-Zumino-Novikov-Witten models (they are those with level $k=1,2,4$), which are however not dense in the Hilbert space of physical states \cite{rev}; this  constitutes a fundamental limitation. 


\paragraph*{3. Alternative realizations of  anyonic theories}

To address the problem, we make the important remark that the Pfaffian state described in the previous paragraph can also be realized in other systems, physically different from the one provided by FQHE. Therefore, it may be possible to realize the same non-local topological excitations, while circumventing the representation-theory complication. To begin, note that the effective quantum theory corresponding to anyonic excitations with $\kappa=4$ is the $\widehat{sl}(2)_{k=2}$ WZNW theory \cite{Bravy}.  


Fixing a $level$ $k \in \mathbb{N}$ of the representation gives the central charge of the theory as $c = 3k/(k+2)$, and leads to a number of possible types of physical excitations, whose commutation relation is of anyonic type, with fractional statistics. Moreover,  the primary field $g(z, \bar z)$ satisfies the equations 
(Knizhnik-Zamolodchikov) 
$$
[(k+2)\p_{z}-J^a\tau^a] g(z, \bar z) = 0, \,\, J^a(\zeta) g(z, \bar z) =\frac{\tau^a}{\zeta-z} g(z, \bar z),
$$
with $\tau^a$ the generators of the Lie algebra $su(2)$. 

It is relevant to note the correspondence between the Kac-Moody algebras $\widehat{sl}(2)_k$ and the quantum group $\mathcal{U}_q(sl(2))$, where the quantization parameter $q$ is a root of unity, $q = \exp [i\pi/(k+2)]$. This  can be regarded as a manifestation of the Riemann-Hilbert correspondence (between monodromy of solutions of linear differential equations and the singular points of the equations). Algebraically, this correspondence identifies the centralizer of $\widehat{sl}(2)_k$ to the Hecke algebra $H(q)$, a $q-$deformation of the symmetric group, and providing representations for the braid group in the corresponding number of elements. 


\paragraph*{4. Cold Fermi gases in the fast-pairing limit}

The physical system we propose to use instead of quantum Hall devices consists of cold 
fermionic atoms, trapped in an optical lattice, and driven to create Cooper pairs through 
the Feshbach resonance \cite{Fermi1}--\cite{Fermi8}.  
It is well-known \cite{Sierra}--\cite{AFS} that a proper 
model for the dynamics of this system is the Richardson-Gaudin pairing model \cite{Richardson1}--\cite{Gaudin76}, which we 
briefly review here. 
It describes a system of $n$ fermions characterized by a set of independent one-particle states of energies $\epsilon_l, l \in \Lambda$. Each state $l$ has a total (spin) degeneracy $d_l=2$, and the states within the subspace corresponding to  $l$ are further labeled by an internal quantum number, $ s= \uparrow, \downarrow $. Let $\hat c^{\dag}_{ls}$ represent the fermionic creation operator for the state $(ls)$. Using the Anderson pseudo-spin operators \cite{Anderson}  (quadratic pairing operators), satisfying the $su(2)$ algebra
\beq
[t^3_i, \, t_j^{\pm}] = \pm \delta_{ij} t_j^{\pm}, \,\,\,\, [t^+_i, \, t^-_j] = 2\delta_{ij}t^3_j,
\eeq
the Richardson pairing Hamiltonian is given by
\beq \label{pairing}
H_P = 
\sum_{l \in \Lambda} 2\epsilon_l t^3_l - g\sum_{l, l'}t^+_l t^-_{l'} = 
\sum_{l \in \Lambda} 2\epsilon_l t^3_l - g{\bf{t}}^+\cdot {\bf{t}}^-, 
\eeq
where ${\bf {t}} = \sum_{l}{\bf {t}}_l$ is the total spin operator. 
It maps to the  reduced BCS model 
\beq \label{Hamiltonian}
\hat H = \sum_{{\bf{p}}, \sigma}\epsilon_{\bf{p}}\hat{c}_{{\bf{p}}, \sigma}^{\dag}\hat{c}_{{\bf{p}}, \sigma}
-g\sum_{{\bf{p}}, {\bf{k}}}\hat{c}_{{\bf{p}} \, \uparrow}^{\dag}\hat{c}_{{-\bf{p}}\, \downarrow}^{\dag}
\hat{c}_{{-\bf{k}}\, \downarrow}\hat{c}_{{\bf{k}}\, \uparrow}
\eeq
by replacing the translational degrees of freedom by rotational ones, where $l  \in \Lambda = \{ 1, \ldots,  n \}$ ennumerates the one-particle orbital degrees of freedom, while $s = \, \uparrow, \, \downarrow$ indicates the two internal spin states per orbital ($d_l=2$). The pairing Hamiltonian can be decomposed into the linear combination
\beq
H_P=2\sum_{l \in \Lambda} \epsilon_l R_l + g \left [ \left ( \sum_{l \in \Lambda} t^3_l \right )^2 - \frac{1}{4} \sum_{l \in \Lambda} (d_l^2-1) \right ].
\eeq
At a fixed value of the component $t^3$ of the total angular momentum, the last term becomes a constant and is dropped from the 
Hamiltonian. The operators $R_l$ (generalized Gaudin magnets \cite{Gaudin76}) are given by
\beq \label{gaudin}
R_l = t^3_l - \frac{g}{2}\sum_{l' \ne l}\frac{{\bf {t}}_l \cdot {\bf {t}}_{l'}}{\epsilon_l - \epsilon_{l'}}.
\eeq
These operators solve the Richardson pairing Hamiltonian because \cite{CRS97} they are independent, commute with
each other, and span all the degrees of freedom of the system. Richardson showed \cite{Richardson1,Richardson2} that the exact $N-$pair wavefunction 
of his Hamiltonian is given by application of operators 
$
b^{\dag}_{k} = \sum_{l}\frac{t_l^{\dag}}{2\epsilon_l - e_k}
$
to vacuum (zero pairs state). The unnormalized $N-$pair wavefunction reads
$
\Psi_R(\epsilon_i) = \prod_{k=1}^Nb_k^{\dag}|0\rangle.
$
The eigenvalues $e_k$ satisfy the self-consistent algebraic equations 
\beq \la{ec}
\frac{1}{g} = \sum_{p \ne k}\frac{2}{e_k - e_p} +\sum_l \frac{1}{2\epsilon_l-e_k}.
\eeq
As an integrable model, the Richardson-Gaudin system was solved based on the methods of \cite{SOV}, in \cite{Kuznetsov}--\cite{Harnad}.


\paragraph*{5. Richardson-Gaudin model as singular limit of Wess-Zumino-Witten theory}

In this section we recall the relation between the Richardson-Gaudin model and
singular $SU(2)$ Chern-Simons and WZW models \cite{Gawedzki1,Gawedzki2}. These relations stem from the study of quantum states of the Chern-Simons theory with gauge group $SU(2)$ on the manifold $\mathbb{T}^2 \times \mathbb{R}$ and in the presence of Wilson lines $\{ z_i\} \times \mathbb{R}$, $i = 1, \ldots, n$ \cite{Gawedzki1}, corresponding to the one-particle spectrum of the Richardson Hamiltonian,
$z_i = 2\epsilon_i$. The torus $\mathbb{T}^2$ has modular parameter $\tau = \tau_1 + i\tau_2, \,\, \tau_{1,2} \in \mathbb{R}$. Thus, $\mathbb{T}^2 = \mathbb{C}/\mathbb{Z} + \tau\mathbb{Z}$. The quantum states of this theory are known to satisfy the Knizhnik-Zamolodchikov-Bernard (KZB) equations 
\beq \label{kzb}
\nabla \Psi_{CS} = 0, \quad \nabla = (\nabla_{\tau}, \nabla_{z_i}), 
\eeq
where $\nabla$ is the flat KZB connection,
\beq \label{kzbcon}
\nabla_{\tau} = (k+2) \p_{\tau} + H_0(z_i), \,
\nabla_{z_i} = (k+2) \p_{z_i} + H_i(z_i), 
\eeq
and $H_0, H_i$ are elliptic versions of the Gaudin Hamiltonia (and depend on the torus parameter $\tau$). The parameter 
$k$ is related to the level of representation of the affine algebra $\widehat{sl}(2)_k$. By a limiting procedure, the system degenerates into the Richardson-Gaudin (R-G) problem: by taking the limit 
$\tau \to i\infty$, the torus degenerates into a cylinder, and the elliptic Hamiltonia $H_i, H_0$ reduce to a trigonometric limit. Upon rescaling such that the set $\{z_i\}$ is collapsed into the origin, the cylinder becomes the complex plane with punctures at $\{z_i\}$, and the rescaled operators become the rational Gaudin magnets (\ref{gaudin}). 

The R-G model is formally retrieved in the singular limit $k+2 \to 0$, which imposes a careful analysis of the semiclassical solution in order to identify the proper choice for the fractional levels $k$ as $c \to -\infty$. This analysis will be carried out in greater detail elsewhere, and we discuss here only the main conclusions. Choosing the sequence of fractional levels $k = 2\frac{1-8m}{1+8m}, m \in \mathbb{N}$, or equivalently $k+2 = \frac{4}{8m+1}, c = \frac{3}{2}(1-8m), q = e^{i\pi/4}$,  the corresponding representation of the braid group $H(q)$ remains fixed as $m \to \infty$. The fractional levels are {\emph{admissible}} in the sense of Kac and Kazhdan \cite{KK}, and they are understood in the sense of \cite{RC}. This generalized notion of representation level settles in particular the issue of irreducible representations of the modular group $SL(2, \mathbb{Z})$. As noted in \cite{RC}, the initial interpretation of fractional levels for $\widehat{sl}(2)$ \cite{KW} led to a {\emph{finite}} number of irreducible representations of $SL(2, \mathbb{Z})$, which is clearly insufficient, as the set of automorphic forms on the Poincar\'e upper half-plane $\mathbb{H}$, invariant under $SL(2, \mathbb{Z})$, of integer weight in $2 \mathbb{N}$ is described by the union of the Berezin kernels from the tower of von Neumann algebras $B(H_n), n \ge 1$, where $H_n  = L^2(\mathbb{H}, y^{n-2} dxdy)$. In contrast to the original Kac-Wakimoto approach to fractional level representations of $\widehat{sl}(2)$ (which leads to a finite number of irreducible representations of the modular group), the generalization \cite{Gaberdiel} provides an infinite number.


\paragraph*{6. Semiclassical solution of the Richardson-Gaudin model}

The solution starts from the Lax operator 
\beq \label{lax} 
\mathcal{L}(\lambda) = \frac{2}{g}\sigma_3 +\sum_{i=0}^n \frac{\vec S_i \cdot\vec \sigma}{\lambda - \epsilon_i} = 
\left [
\begin{array}{cr}
a(\lambda) & \,\, \, b(\lambda) \\
c(\lambda) & -a(\lambda)
\end{array}
\right ],
\eeq
where $\sigma_\alpha, \, \alpha = 1,2,3$ are the Pauli matrices, and $\lambda$ is an additional complex variable, the spectral 
parameter. Let $u_k, \, k=1, \ldots, n-1$ be the roots of the coefficient $c(\lambda)$. 
Poisson brackets for variables $S_i^{\alpha}$ read 
\beq
\{S_j^\alpha , \, S_k^{\beta} \}=2\epsilon_{\alpha \beta \gamma}S_k^\gamma \delta_{jk}
\eeq
The Lax operator (\ref{lax}) defines 
a Riemann surface (the spectral curve) $\Gamma (y, \lambda)$ of genus 
$g = n-1$, through 
\beq \label{curve}
y^2 = Q(\lambda) = \det \mathcal{L}(\lambda)\left [ g\frac{P(\lambda)}{2} \right ]^2 , 
\eeq
where $P(\lambda) = \prod_{i=1}^n (\lambda -  \epsilon_i).$

The equations of motion become
$$ 
\dot  u_i = \frac{2i y( u_i)}{\prod_{j \ne i}( u_i - u_j)},
\quad
i\dot J^- = J^{-} \left [ gJ^3 + 2\sum_{k=1}^{n} \epsilon_k - u \right ].
$$
Here, 
$u = -2\sum_{i=1}^{n-1}  u_i, \quad b(u_i) = 0$. 

From the equations of motion, it is clear that knowledge of the initial amplitude of $J^-$ and of the roots $ \{  u_i \}$
is enough to specify the $n$ unit vectors $\{ {\bf {S}}_i \},$ for a given set of constants of motion $\{ R_l \}$ given by the classical limit of Gaudin Hamiltonia. The Dubrovin equations of motion are 
solved by the inverse of the Abel-Jacobi map, as we explain in the following. We begin by noting that the polynomial 
$Q(\lambda)$ has degree $2n$, and is positively defined on the real $\lambda$ axis. Therefore, the curve $\Gamma (y, \lambda)$
has $n$ cuts between the pairs of complex roots $[E_{2i-1}, \, E_{2i}], i = 1, 2, \ldots, n$, perpendicular to the real $\lambda$
axis. The points $u_i$ belong to $n-1$ of these cuts, $u_i \in [E_{2i-1}, \, E_{2i}], i =1, \ldots, n-1$. These 
$g = n-1$ cuts allow to define a canonical homology basis of $\Gamma$, consisting of cycles $\{\alpha_i, \beta_i \}, i = 1, \ldots, g$. With respect to these cycles, a basis of normalized holomorphic differentials $\{ \omega_i \}$ can be defined, through
\beq
\mu_i = \lambda^{g-i}\frac{d\lambda}{y}, \,\,\, M_{ij} = \int_{\alpha_j} \mu_i, \,\,\, {\bm {\omega}} = M^{-1}{\bm{\mu}}. 
\eeq 
The period matrix $B_{ij} = \int_{\beta_j} \omega_i$
is symmetric and has positively defined imaginary part. The Riemann $\theta$ function is defined with the help of the period matrix as
\beq
\theta({\bm{z}} | B) = \sum_{{\bm{n}} \in {\bm{z}}^g} e^{2\pi i ({\bm {n}}^t {\bm{z}} + \frac{1}{2} {\bm {n}}^t B {\bm {n}} )}.
\eeq
The $g$ vectors $\bm{B}_k$ consisting of columns of $B$ and the basic
vectors $\bm{e}_k$  define a lattice in ${\mathbb{C}}^g$. The $Jacobian$ variety of the curve $\Gamma$, is then the $g-$dimensional torus defined as the quotien 
$J(\Gamma)  = \mathbb{C}^g /(\mathbb{Z}^g + B\mathbb{Z}^g).$
The Abel-Jacobi map associates to any point $P$ on $\Gamma$, a point ($g-$ dimensional complex vector) on the Jacobian variety, through
${\bm {A}} (P) = \int_{\infty}^P {\bm {\omega}}.$
Considering now a $g-$dimensional complex vector of points $\{P_k \}, 
k = 1, \ldots, g$ on $\Gamma$, defined up to a permutation, we can associate to it the point on the Jacobian
\beq  \label{a}
{\bm{z}} = {\bm{a}}({\bm {P}}) = \sum_{k=1}^g {\bm {A}} (P_k) + {\bm {K}},
\eeq
where ${\bm {K}}$ is the Riemann characteristic vector for $\Gamma$. 

The map (\ref{a}) suggests that we now have a 
way to describe the dynamics on $\Gamma$ by following the image point on the Jacobian. 
Given a point on the $g-$dimensional Jacobian  ${\bm{z}} = (\zeta_1, \ldots , \zeta_{n-1})$, we can find an unique set of points $\{ \lambda_k\}, k = 1, \ldots, g$ on $\Gamma$, such that ${\bm{z}} = {\bm {a}}(
{\bm {\lambda}}),$ and $\theta ({\bm{a}(\bm{P}) - \bm{z}} | B) = 0$. The system evolves in time according to the point ${\bm{z}}(t)$ 
\beq \label{solution1}
\zeta_k = ic_k, \,\, 1 \le i \le g-1, \,\, \zeta_{n-1} = i(c_{n-1} + t),
\eeq
where $\{ c_k\}$ is a set of initial conditions, such that  $
{\bm{z}}_0 = {\bm{z}}(t=0) = {\bm {a}}({\bm{c}})$, and ${\bm{c}}$ is the set of initial
conditions for positions of ${\bm {\lambda}}$ on $\Gamma$. Together with the initial condition which determines the
initial amplitude of $J^-$, this set completely determines the functions $ u_i(t), J^-(t)$. The application of inter-twiners of the Hecke algebra $H(q)$ will be reflected in the homological structure of the Sklyanin solution for the semiclassical R-G problem as the coalescence of a pair of branch points (followed by their subsequent resolution), which can be translated directly at the level of the solution $J^{-}(t)$. 



%

\end{document}